\documentclass[aps,pra,twocolumn,showpacs,preprintnumbers,amsmath,amssymb,footinbib]{revtex4}
\usepackage{mathtools}

\DeclarePairedDelimiter{\floor}{\lfloor}{\rfloor}
\usepackage{graphicx,epsfig}
\usepackage{bm}
\usepackage{dcolumn}
\usepackage{xcolor}% coloured text
\usepackage[breaklinks=true,colorlinks,citecolor=blue,linkcolor=blue,urlcolor=blue]{hyperref}

\def\nat#1#2#3{Nature {\bf #1}, #2 (#3)}
\def\np#1#2#3{Nat. Phys. {\bf #1}, #2 (#3)}

\def\prv#1#2#3{Phys. Rev. {\bf #1}, #2 (#3)}
\def\rmp#1#2#3{Rev. Mod. Phys. {\bf #1}, #2 (#3)}
\def\prl#1#2#3{Phys. Rev. Lett. {\bf #1}, #2 (#3)}
\def\pra#1#2#3{Phys. Rev. A {\bf #1}, #2 (#3)}
\def\prb#1#2#3{Phys. Rev. B {\bf #1}, #2 (#3)}

\def\jpamt#1#2#3{J. Phys. A: Math. Theor. {\bf #1}, #2 (#3)}

\def\ajp#1#2#3{Am. J. Phys. {\bf #1}, #2 (#3)}

\def\jcp#1#2#3{J. Chem. Phys. {\bf #1}, #2 (#3)}

\def\noi{\noindent}
\def\bc{\begin{center}}
\def\ec{\end{center}}
\newcommand{\bea}{\begin{equation}}
\newcommand{\eea}{\end{equation}\noi}
\newcommand{\ber}{\begin{eqnarray}}
\newcommand{\eer}{\end{eqnarray}\noi}

\begin{document}
\title{Re-examining Einstein's $B$ coefficient and rate equations with the Rabi model}
%\author{Tanmoy Mondal}
%\author{Shyamal Biswas}\email{sbsp [at] uohyd.ac.in}
%\affiliation{School of Physics, University of Hyderabad, C.R. Rao Road, Gachibowli, Hyderabad-500046, India}

\author{Najirul Islam$^{1}$}
\author{Tanmoy Mondal$^{1}$\footnote{Present Address: Harish-Chandra Research Institute, Chhatnag Road, Jhusi, Allahabad-211019, India}}
\author{Sagar Chakraborty$^{2}$}
\author{Shyamal Biswas$^{1}$}\email{sbsp [at] uohyd.ac.in}

\affiliation{$^{1}$School of Physics, University of Hyderabad, C.R. Rao Road, Gachibowli, Hyderabad-500046, India\\
$^{2}$Department of Physics, Indian Institute of Technology Kanpur, Kanpur-208016, India}

\date{\today}

\begin{abstract}
Starting from the Rabi Hamiltonian, which is useful in arriving at non-perturbative results within the rotating wave approximation, we have found Einstein's $B$ coefficient to be time-dependent: $B(t)\propto|J_0(\omega_\gamma t)|$ for a two-level system (atom or molecule) in thermal radiation field. Here $\omega_\gamma$ is the corresponding Rabi flopping (angular) frequency and $J_0$ is the zeroth order Bessel function of the first kind. The resulting oscillations in the $B$ coefficient---even for very small $\omega_\gamma$---drives the system away from thermodynamic equilibrium at any finite temperature contrary to Einstein's assumption. The time-dependent generalized $B$ coefficient facilitates a path to go beyond Pauli's formalism of non-equilibrium statistical mechanics involving the quantum statistical Boltzmann (master) equation. In this context, we have obtained entropy production of the two-level system by revising Einstein's rate equations, while considering the $A$ coefficient to be the original time-independent one and the $B$ coefficient to be time-dependent.
\end{abstract}

\pacs{03.65.Sq, 05.30.-d, 03.65.-w}

\maketitle
 
%\tableofcontents

\section{Introduction}
Einstein's $A$ and $B$ coefficients are quite known to the scientific community in connection with the formation of spectral lines involving fundamental processes, such as spontaneous emission, stimulated absorption and stimulated emission, undergone on a two-level system (atom or molecule) in the presence of an oscillatory electromagnetic field, say laser light, thermal radiation, \textit{etc.}~\cite{Einstein}. While the $A$ coefficient is the rate of spontaneous emission from a higher energy level to a lower energy level of the two-level system caused by vacuum fluctuations of electromagnetic field, the $B$ coefficient is the rate of stimulated absorption $B_{12}$ (or emission $B_{21}$) of (or from) the same system in the radiation field for unit energy density of the radiation per unit (angular) frequency interval around the Bohr frequency~\cite{Einstein,Hilborn}. Einstein's $A$ and $B$ coefficients are of very high importance, as because, the spectral lines have huge applications almost everywhere in the modern science, engineering, and technology. These coefficients also determine density of photons in thermal equilibrium when the probability of transitions for a two-level system reaches a steady state.

Historically, almost a century back---during the era of old quantum mechanics---when time-dependent perturbation theory was not known~\cite{Einstein}, Einstein's $A$ and $B$ coefficients were proposed to be time-independent. These coefficients, for the two-level system in thermal radiation field at an absolute temperature $T$, were determined in terms of fundamental constants by Dirac, Weisskopf, and Wigner in the quantum mechanics era within (i) the frameworks of the time-dependent perturbation theory for the light-matter interactions and (ii) the quantum field theory of the stimulated emission, the stimulated absorption, and the spontaneous emission of radiation~\cite{Dirac,Weisskopf}. However, neither Einstein's semi-classical theory of radiation~\cite{Einstein} nor Dirac's first order quantum mechanical perturbation theory of radiation~\cite{Dirac} predicted regularity in the stimulated transitions (absorption and emission) though the electromagnetic field incident on the two-level system oscillates in a regular manner. This regularity was predicted a decade later by Rabi~\cite{Rabi}. He and his collaborators showed resonance in the two-level system in the course of stimulated absorption and emission within a nonperturbative model which is now known as the Rabi model~\cite{Rabi,Rabi2}. For the two-level system (having the electric dipole moment $\vec{d}$ and the Bohr (angular) frequency $\omega_0$ corresponding to energy eigenstates $|{\psi_1}\rangle$ and $|\psi_2\rangle$) in presence of an oscillatory electromagnetic field with the electric field component $\vec{E}=\vec{E_0}\cos(\omega t)$, Rabi \textit{et al.} obtained a generalized (angular) frequency for the transition induced flopping of the two states, as $\Omega=\sqrt{(\omega-\omega_0)^2+\omega_\gamma^2}$ which is now known as the Rabi formula where $\omega_\gamma=|\langle\psi_1|\vec{d}\cdot \vec{E}_0|\psi_2\rangle|/\hbar$ is the Rabi flopping frequency~\cite{Rabi2}. 

It it quite known, that, the generalized Rabi flopping frequency ($\Omega$) tends to the Rabi flopping frequency ($\omega_\gamma$) at resonance ($\omega\rightarrow\omega_0$), i.e., where the probability of the stimulated transitions  for the stimulated emission from the initial ($t=0$) state $|\psi_2\rangle$ to the final state $|\psi_1\rangle$ at time $t$, say $P_{2\rightarrow1}(t)=\omega_\gamma^2\frac{\sin^2(\Omega t/2)}{\Omega^2}=\omega_\gamma^2\frac{\sin^2(\sqrt{(\omega-\omega_0)^2+\omega_\gamma^2}t/2)}{(\omega-\omega_0)^2+\omega_\gamma^2}$, is sharply peaked~\cite{Rabi2,Griffiths,Sakurai,Shirley}.  The expression of transition probability (involving the Rabi flopping frequency) is quite successful, as it gives a reliable value of the nuclear magnetic moment to experimentalists~\cite{Rabi2}. Later experimentalists found this expression quite successful for atoms, molecules, semiconductors, Bose--Einstein condensates, many-bodies, \textit{etc.} exposed in laser light~\cite{Hocker,Cundiff,Brune,Meekhof,Donley,Udem,Dudin}. One can get the perturbation result (which is compatible with Fermi's golden rule)~\cite{Dirac,Griffiths} back if one assumes $|\omega-\omega_0|\gg\omega_\gamma$ in the Rabi formula. But, condition for the time-dependent perturbation ($P_{2\rightarrow1}(t)\ll1~\forall~t$, i.e., $\omega_\gamma^2/(\omega-\omega_0)^2\ll1$) does not hold~\cite{Griffiths} at the resonance ($\omega\rightarrow\omega_0$) at least for $t\rightarrow\infty$ however weak the light-matter coupling ($\hbar\omega_\gamma$) may be. The problem with the upper limit of time ($0\le t<\infty$) can not be avoided to get the frequency matching condition ($\delta(\omega-\omega_0)$~\footnote{The frequency matching condition follows from the limiting case of the square root of the stimulated transition probability $\lim_{t\rightarrow\infty}\sqrt{P_{2\rightarrow1}(t)}=\lim_{t\rightarrow\infty}\omega_\gamma\frac{\sin([\omega-\omega_0]t/2)}{\omega-\omega_0}=\omega_\gamma2\pi\delta(\omega-\omega_0)$ within the 1st order sinusoidal perturbation.}) for the dipole-transitions stimulated by a sinusoidal perturbation~\cite{Dirac,Griffiths}. Thus, the divergence of the transition probability questions soundness of the 1st order perturbation theory at the resonance for $t\rightarrow\infty$. The soundness can, of course, be restored only for $\omega_\gamma\rightarrow0$ so that $\lim_{\omega_\gamma\rightarrow0, \omega\rightarrow\omega_0}\omega_\gamma\delta(\omega-\omega_0)=\text{constant}\lnsim1$.

Since the condition for the sinusoidal perturbation with non-vanishing light-matter coupling ($\hbar\omega_\gamma$) is not satisfied~\cite{Griffiths} at the resonance ($\omega\rightarrow\omega_0$) where the stimulated transitions (emission) are most probable, the first order perturbation result ($B_{12}=B_{21}=\frac{\pi}{3\epsilon_0\hbar^2}|\langle\psi_1|\vec{d}|\psi_2\rangle|^2$~\cite{Dirac,Griffiths}~\footnote{If degeneracy of the two states are $g_1$ and $g_2$, respectively, then $B_{21}/B_{12}$ would be given by $B_{21}/B_{12}=g_1/g_2$~\cite{Hilborn}.}) for the $B$ coefficient obtained by Dirac~\cite{Dirac} is not reliable for $\omega_\gamma\nrightarrow0$. This is a problem with the quantum mechanical perturbation theory, and it remains so even in the weak coupling regime ($\omega_\gamma/A\ll1$) as long as the coupling ($\hbar\omega_\gamma$) does not tend to zero keeping the natural decay rate (i.e., the $A$ coefficient) fixed to a nonzero value. Consequently, a question arises: what would be the reliable expression for the $B$ coefficient in the weak coupling regime? This issue could have been addressed with a nonperturbative model such as the Rabi model~\cite{Rabi,Rabi2}, but has surprisingly been overlooked for the last eight decades, though there are several works done in the intermediate regime ($0\lnsim\omega_\gamma/A\lesssim1$) with the consideration of moderate system-bath (i.e., system-radiation field) interactions. These interactions are often (i) semiclassically modelled as perturbation terms in the Bloch--Redfield (Markovian master) equation within the density matrix formalism~\cite{Wollfarth,Leggett}, (ii) quantum mechanically modelled as non-perturbation terms in the Schrodinger equation within the generalized Weisskopf-Wigner (natural) decay formalism for discrete and continuum modes~\cite{Cohen-Tannudji2}, and (iii) quantum mechanically modelled as non-perturbation terms in the Nakajima--Zwanzig-type (non-Markovian master) equation within the density matrix formalism~\cite{Nakajima,Anastopoulos}.

While the stimulated transition-rates are time-independent in the (quantum) Markovian master equations and are solvable for the case of the time-dependent perturbation on the system~\cite{Lindblad,Leggett,Wollfarth}, they are time-dependent in the (quantum) non-Markovian master equations and are usually hard to solve~\cite{Nakajima,Anastopoulos}; the explicit time-dependent terms in the stimulated transition rates render the non-Markovian master equations analytically intractable. Of course, some simplified versions of the non-Markovian master equations can be solved either in the limiting cases of weak~\cite{Intravaia} and linear~\cite{Chakraborty} interactions between the system and the bath or in the limiting case of structured bath even for strong interactions~\cite{Escher}.  Nevertheless, population dynamics of the open quantum system of our interest, i.e., the two-level system in the thermal radiation field, has not been described so far through exact analytical solutions of the non-Markovian master equations.

The semiclassical Rabi model, we are considering, though is a non-perturbative one, gives exact results even in the weak coupling regime, as the model is exactly solvable for the entire range of coupling constant ($\hbar\omega_\gamma$). Hence, we aim (i) to get a single reliable expression of Einstein's $B$ coefficient from the (semiclassial) Rabi model not only for the weak coupling regime ($\omega_\gamma/A\ll1$) but also for the entire regime including the strong coupling regime ($\omega_\gamma/A\gg1$), (ii) to generalize Einstein's rate equations with the reliable $B$ coefficient for the two-level system in the thermal radiation field, and (iii) to describe population dynamics of the system by finding exact analytic solutions of the rate equations. It should be mentioned in this regard that, the thermal radiation field is not coherent. The population dynamics of the two-level system in the coherent radiation field is also of high interest, and has been experimentally investigated in a high-$Q$ cavity by Brune \textit{et al.}~\cite{Brune}. Theoretical explanation of the same has been found numerically by Escher and Ankerhold~\cite{Escher}, on the dissipative quantum dynamics of the two-level system interacting with a structured reservoir consisting of damped harmonic modes. However, we are aiming at the light-matter interactions at the semi-classical level~\footnote{Here in the semiclassical theory, while both the system and the bath (i.e., the thermal radiation field of photons) are treated quantum mechanically, the system-bath interactions are treated classically by not considering annihilation and creation operators in the interactions} (and not at the quantum mechanical level) so that the population dynamics can be analytically described in a fairly accurately manner, at least for $\omega_0\gg\omega_\gamma$ and $k_BT\gnsim \hbar\omega_\gamma$.

The next section of this paper begins with the Rabi model for the two-level system in a sinusoidally oscillating electromagnetic field~\cite{Rabi,Rabi2,Griffiths,Sakurai}. Then we write down transition probabilities for the electric-dipole transitions among the two (energy) levels, and recast the transition probabilities for the two-level system in the thermal radiation by integrating over all possible frequencies and polarizations of the thermal radiation field. This result significantly differs from the perturbation result. Here, we have considered zero-point energy of the thermal radiation field in all our analyses to get generalized semiclassial results on top of the semiclassical theory~\footnote{Semiclassical results, in this respect, are found with quantum mechanical treatment of the system and semi-classical treatment of the light-matter interactions involving statistics of photons without considering its operator algebra in the course of absorption and emission~\cite{Griffiths,Sakurai}. Inclusion of the operator algebra for photon-annihilation and photon-creation operators would make the treatment quantum electrodynamic (QED)~\cite{Jaynes}.}. We also have obtained generalized semiclassical result for vacuum Rabi flopping of the two-level system with the consideration of small contribution of the thermal photons in a resonant cavity at a low temperature. We have compared this result with the experimental data obtained by Brune \textit{et al.}~\cite{Brune}. We get oscillatory type time-dependent $B$ coefficient, from the corresponding transition-probability (i.e., the stimulated emission's probability) for the thermal photons. Then we revise Einstein's rate (master) equations, considering the $A$ coefficient to be the original time-independent one and the $B$ coefficient to be the time-dependent one, to get the time evolution of the occupation probabilities of the two levels~\cite{Einstein}. We get entropy production of the two-level system, from these time-dependent probabilities, by following Pauli's formalism of nonequilibrium statistical mechanics~\cite{Pauli,Feynman}. We have also considered monochromatic radiation field side by side throughout the paper. Finally, we discuss our results.
\section{Two-level system in thermal radiation field}
\subsection{Rabi model}
The Rabi Hamiltonian for the two-level system having electric dipole moment $\vec{d}$ in the oscillatory electromagnetic field (with the electric field component $\vec{E}=\vec{E}_0\cos(\omega t)$) is given by~\cite{Rabi,Rabi2,Sakurai,Cohen-Tannudji}
\begin{eqnarray}\label{eqn:1}
H&=&E_1|\psi_1\rangle\langle\psi_1|+E_2|\psi_2\rangle\langle\psi_2|\nonumber\\&&-\frac{\vec{E}_0\cdot\vec{d}}{2}\big[e^{i\omega t}|\psi_1\rangle\langle\psi_2|+e^{-i\omega t}|\psi_2\rangle\langle\psi_1|\big],~~~
\end{eqnarray}
where $|\psi_1\rangle$ and $|\psi_2\rangle$ constitute a set of two orthonormal states of the two-sate system (in absence of the external field) with energy eigenvalues $E_1$ and $E_2$ ($E_2>E_1$) respectively. The third term of the Hamiltonian represents the semiclassical interaction between the two-level system (atom or molecule) and the external electromagnetic field. The interaction term, although is not a perturbation, is consistent with the rotating wave approximation ($\omega_0+\omega\gg|\omega_0-\omega|$) which is also used in the time-dependent perturbation theory~\cite{Rabi2,Griffiths}. Validity of the rotating wave approximation, however, is not questioned at the resonance ($\omega\rightarrow\omega_0$) as long as $\omega_0=(E_2-E_1)/\hbar$ is fairly large, say $\omega_0\gg\omega_\gamma$. Thus, the Rabi model is applicable for large Bohr frequency ($\omega_0$) of the two-level system, and incidentally, the Schrodinger equation for the two-level system corresponding to the Rabi Hamiltonian is exactly solvable under the transformation into the interaction-picture~\cite{Rabi2,Griffiths}. The Rabi model is of course an integrable one due to the presence of a discrete symmetry in it~\cite{Braak}. The energy eigenvalues of the Rabi Hamiltonian in Eq.~(\ref{eqn:1}) thus takes the form $E_\mp=[E_2+E_1]/2\mp\frac{\hbar}{2}\sqrt{(\omega-\omega_0)^2+\omega_\gamma^2}$~\cite{Rabi2}. Corresponding eigenstates are now dressed due to the light-matter coupling resulting the energy eigenvalues different from those ($E_1, E_2$) of the uncoupled bare states ($|\psi_1\rangle, |\psi_2\rangle$). Both the eigenstates eventually are linear combinations of the uncoupled bare states: $|\psi_-\rangle=\cos(\theta)|\psi_1\rangle+\sin(\theta)|\psi_2\rangle$ \& $|\psi_+\rangle=-\sin(\theta)|\psi_1\rangle+\cos(\theta)|\psi_2\rangle$ for $\tan(\theta)=\omega_\gamma/[\sqrt{(\omega-\omega_0)^2+\omega_\gamma^2}-(\omega-\omega_0)]$~\cite{Benson}.

Eventually, the two-level system will always be in the superposition state $|\psi\rangle$ of the two energy eigenstates as well as of the uncoupled bare states as long as energy of the system is not measured. Thus, time evolution of the state of the system takes the following form~\footnote{It takes another form, viz., $|\psi(t)\rangle=c_-e^{-iE_-t/\hbar}|\psi_-\rangle+c_+e^{-iE_+t/\hbar}|\psi_+\rangle$, in the basis of the energy eigenstates with time-independent coefficients ($c_\mp$) resulting in no transitions between dressed eigenstates $|\psi_+\rangle$ and $|\psi_-\rangle$.}:  $|\psi(t)\rangle=c_{1}(t)e^{-iE_1t/\hbar}|\psi_1\rangle+c_{2}(t)e^{-iE_2t/\hbar}|\psi_2\rangle$, where $c_1(t)$ is the transition probability amplitude for the dipole-transition from the uncoupled bare state $|\psi_2\rangle$ to $|\psi_1\rangle$ and $c_2(t)$ is that from the uncoupled bare state $|\psi_1\rangle$ to $|\psi_2\rangle$~\cite{Dirac,Rabi,Rabi2}. Transformation of the Schrodinger equation for the two-level system corresponding to the Rabi Hamiltonian into the interaction-picture, results in the transition probability,
\begin{eqnarray}\label{eqn:2}
P_{2\rightarrow1}(t)=|c_1(t)|^2=\omega_\gamma^2\frac{\sin^2(\Omega t/2)}{\Omega^2},
\end{eqnarray}
on using the suitable boundary condition that the system was initially ($t=0$) only in the state $|\psi_2\rangle$~\cite{Rabi2,Griffiths,Sakurai}. Here, $\Omega=\sqrt{(\omega-\omega_0)^2+\omega_\gamma^2}=(\omega-\omega_0)\sqrt{1+[\frac{\omega_\gamma}{\omega-\omega_0}]^2}$ is the generalized Rabi flopping (angular) frequency and $\omega_\gamma=|\langle\psi_1|\vec{d}\cdot \vec{E}_0|\psi_2\rangle|/\hbar$ is the Rabi flopping frequency for the stimulated emission from the state $|\psi_2\rangle$ to the state $|\psi_1\rangle$. Also, we keep in mind that
\begin{eqnarray}\label{eqn:3}
P_{1\rightarrow2}(t)=|c_2(t)|^2 =1-P_{2\rightarrow1}(t),
\end{eqnarray}
is the transition probability for the stimulated absorption from the state $|\psi_1\rangle$ to the state $|\psi_2\rangle$.

Eqs.~(\ref{eqn:2}) and (\ref{eqn:3}) hold good for a linearly polarized monochromatic light (having energy density, $u=\frac{1}{2}\epsilon_0E_0^2$~\footnote{Actual energy density, $u=\epsilon_0E_0^2\cos^2(\omega t)$, where magnetic field part also contributes equally is averaged out here, as because, (i) $\omega$ goes to $\omega_0$ at the resonance, and (ii) electromagnetic field oscillates many times within a single Rabi cycle for $\omega_0\gg\omega_\gamma$~\cite{Griffiths}.}) incident on the two-level system. 

\subsection{Effect of quantum fluctuations and thermal fluctuations in free space}
Let us now consider the system in a thermal radiation field in free space~\footnote{Here free space is an idealization of a big black-body cavity of volume $V\rightarrow\infty$. The two-level system would not come to equilibrium with the thermal radiation field in ideal free space.} where all possible frequencies of the incident light are present with two independent arbitrary polarization directions. While averaging over the polarization states there results in factor $1/3$~\cite{Griffiths} in the r.h.s. of Eq.~(\ref{eqn:2}), contribution of the thermal radiation of all possible frequencies leads to an integration in the r.h.s. of Eq.~(\ref{eqn:2}) over $\omega$ with the weight-factor $u(\omega)$. This follows from Planck's distribution formula (or Bose-Einstein statistics for photons) with adequate corrections for the vacuum energy density. Thus, we recast Eq.~(\ref{eqn:2}), as
\begin{eqnarray}\label{eqn:4}
P_{2\rightarrow1}(t)=\frac{|\langle\psi_1|\vec{d}|\psi_2\rangle|^2}{3\hbar^2}\int_0^\infty\frac{2u(\omega)}{\epsilon_0}\frac{\sin^2(\Omega t/2)}{\Omega^2}d\omega,
\end{eqnarray}
where~\cite{Einstein2,Griffiths}
\begin{eqnarray}\label{eqn:5}
u(\omega)=\frac{\hbar\omega^3}{\pi^2c^3}\bigg[\frac{1}{e^{\hbar\omega/k_BT}-1}+\frac{1}{2}\bigg]
\end{eqnarray}
is the average energy density of the thermal radiation (electromagnetic) field per unit (angular) frequency interval incident on the two-level system~\footnote{Here, time averaging is taken in the very short time scale of $1/\omega_0$. Grand canonical ensemble averaging is further taken for the thermal photons}. While the first term ($\frac{\hbar\omega^3}{\pi^2c^3}\big[\frac{1}{e^{\hbar\omega/k_BT}-1}\big]=u_T(\omega)$, say) in the expression for $u(\omega)$ in Eq.~(\ref{eqn:5}) incidentally represents the average contribution for the thermal photons at the temperature $T$ and is responsible for stimulated emission, the second term ($\frac{\hbar\omega^3}{2\pi^2c^3}=u_q(\omega)$, say) generalizes the semiclassical result and accounts for the correction due to the zero-point energy of the thermal radiation field.

The generalized Rabi frequency ($\Omega$), by the definition, ranges form $-\sqrt{\omega_0^2+\omega_\gamma^2}$ to $-\omega_\gamma$ and $\omega_\gamma$ to $\infty$ following the avoided crossing as $\omega$ varies from $0$ to $\omega_0$ and $\omega_0$ to $\infty$. The integrand in Eq.~(\ref{eqn:4}) is peaked at the resonant frequency, $\omega=\omega_0$ (i.e., at $\Omega=\pm\omega_\gamma$), so that most of the integration comes from $\omega$ close to $\omega_0$.  Further considering $\omega_0$ to be fairly large  (i.e., $\omega_0\gg\omega_\gamma$ which is compatible with the rotating-wave approximation), we recast Eq.~(\ref{eqn:4}), as
\begin{eqnarray}\label{eqn:6}
P_{2\rightarrow1}(t)&\simeq&\frac{2\mu_{12}^2u(\omega_0)}{3\epsilon_0\hbar^2}\bigg[\int_{-\infty}^{-\omega_\gamma}\frac{\sin^2(\Omega t/2)}{\Omega^2}\frac{1}{\sqrt{1-(\omega_\gamma/\Omega)^2}}d\Omega\nonumber\\&&+\int_{\omega_\gamma}^\infty\frac{\sin^2(\Omega t/2)}{\Omega^2}\frac{1}{\sqrt{1-(\omega_\gamma/\Omega)^2}}d\Omega\bigg]\nonumber\\&=&\frac{2\mu_{12}^2u(\omega_0)}{3\epsilon_0\hbar^2}\frac{\pi}{2\omega_\gamma}~_1F_2\big(\{\frac{1}{2}\}, \{1,\frac{3}{2}\}, -\frac{\omega_\gamma^2 t^2}{4}\big)\omega_\gamma t,
\end{eqnarray}
where $_1F_2$ is a generalized hypergeometric function~\footnote{$_1F_2\big(\{1/2\}, \{1, 3/2\}, -(1/4) a^2 x^2\big)=\frac{1}{x}\int J_0(a x)\text{d}x$}, $\mu_{12}=|\langle\psi_1|\vec{d}|\psi_2\rangle|=|\langle\psi_2|\vec{d}|\psi_1\rangle|$ is the transition dipole moment, and the lower limit $\Omega=-\sqrt{\omega_0^2+\omega_\gamma^2}$ (which follows from the avoided crossing) has been replaced~\footnote{Correction to the integration, for this replacement, quickly vanishes as $[-\pi/2 + \text{Si}(\omega_0 t)+\textit{O}(\omega_\gamma^2/\omega_0^2)]t$.} by $-\infty$ as the typical full-width of the transition probability, $\Delta\Omega=4\pi/t$, is well contained (for reasonable values of $t$) within the lowest possible value ($-\sqrt{\omega_0^2+\omega_\gamma^2}$) and the highest possible value ($\infty$) of $\Omega$ for fairly large $\omega_0$. All these approximations for evaluating the above integrations are also applied in the time-dependent perturbative calculation keeping $\omega_\gamma\rightarrow0$~\cite{Dirac,Griffiths}. We differ from the perturbation result~\cite{Dirac,Griffiths} only by keeping $\omega_\gamma\neq0$.

\begin{figure}
\includegraphics[width=0.98 \linewidth]{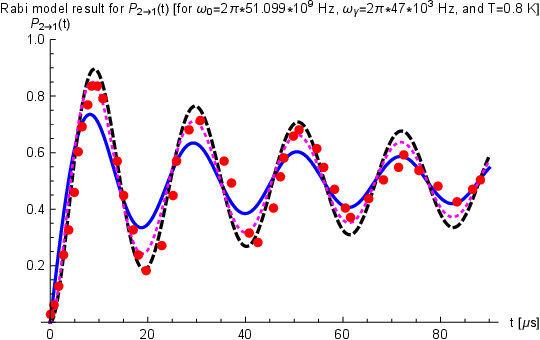}
\caption{
Probability of transitions from the upper level to the lower level of the two-level system in thermal radiation field with the condition that the system initially was in the upper level. Solid line represents free-space Rabi flopping in thermal radiation field, and follows Eq.~(\ref{eqn:6b}) for the parameters as mentioned in the figure. Dashed line follows Eq.~(\ref{eqn:6c}), and represents cavity-space vacuum Rabi flopping  for the natural decay rate $A=0.5536116\times10^6$/s with negligible contribution of the thermal photons in the resonant cavity of $Q$-factor $Q=7\times10^7$. Dotted line represents fitting of the same equation for the same parameters except for the fitted value ($A=1\times10^6$/s) of the natural decay rate. Circles represent experimental data adapted for the circular Rydberg states (with the principal quantum number $n=50$ and $n=51$) of $^{87}$Rb atoms in an open resonant cavity of $Q$-factor $7\times10^7$ and size $\pi(50/2)^2\times27$ mm$^3$ at the temperature $T=0.8$ K~\cite{Brune}.
\label{fig1b}}
\end{figure}

The natural question arises about quantifying the Rabi flopping frequency for the two-level system in the thermal radiation field. Eq.~(\ref{eqn:6}) is the generalization of Eq.~(\ref{eqn:2}) with proper normalization for all the frequencies of the thermal radiation around $\omega_0$. Thus for $t\rightarrow\infty$ and $\omega\rightarrow\omega_0$, the right hand sides of both Eq.~(\ref{eqn:2}) and Eq.~(\ref{eqn:6}) are averaged out to $1/2$ under the consideration that $u(\omega_0)$ remains fixed for emission (including the spontaneous ones) and absorption processes. Thus, we get the Rabi flopping frequency for the two-level system in the thermal radiation field in the free space as
\begin{eqnarray}\label{eqn:6a}
\omega_\gamma=\frac{2\pi\mu_{12}^2u(\omega_0)}{3\epsilon_0\hbar^2}.
\end{eqnarray}
This form of the Rabi flopping frequency is quite general for the two-level system, and is of course, unaltered even for the case of vacuum ($u(\omega_0)=u_q(\omega_0)=\frac{\hbar\omega_0^3}{2\pi^2c^3}$, $u_T(\omega_0)=0$) Rabi flopping~\cite{Jaynes} as $u(\omega_0)$, by definition (Eq.~(\ref{eqn:5})), contains non-zero vacuum energy density per unit frequency interval around $\omega=\omega_0$. Now, using Eqs.~(\ref{eqn:6}) and (\ref{eqn:6a}), we get the probability for the stimulated transitions in free space from the upper level to the lower level, as
\begin{eqnarray}\label{eqn:6b}
P_{2\rightarrow1}(t)=\frac{\omega_\gamma t}{2}~_1F_2\big(\{\frac{1}{2}\}, \{1,\frac{3}{2}\}, -\frac{\omega_\gamma^2 t^2}{4}\big).
\end{eqnarray}
Eq.~(\ref{eqn:6b}) is our generalized semiclassical result for the Rabi flopping in free space. It is called so because non-negligible effect of the zero-point energy of the electromagnetic field has been considered on top of the semiclassical result. We represent this transition probability by the solid line in FIG. \ref{fig1b}. It is clear from the plot that the rate of stimulated transitions is {time-dependent}. It is also clear from this plot that the Rabi flopping is dissipated in the thermal radiation field as expected from the system-bath (i.e., the system-radiation field) interactions that are often modelled perturbatively within the density matrix formalism by the Bloch--Redfield (master) equation~\cite{Wollfarth,Leggett}, in free space. Though additional dissipations are expected from the spontaneous emission (natural decay), our approach is quite non-perturbative, and goes beyond the Bloch--Redfield formalism. Time-averaged {transition} probability was alternatively calculated within the same (semiclassical) Rabi model not for multi-frequency components rather for a single frequency component of the incident electromagnetic field decades back by Shirley~\cite{Shirley}. His result on the {time-averaged} transition probability is significantly different from our net time-dependent transition probability, presented in Eq.~(\ref{eqn:6b}), where contributions of all the frequency components of the thermal radiation field are averaged out with the Planck's distribution.

However, it needs a full quantum electrodynamic (QED) description to capture all the features of the electromagnetic interactions of the two-level system (atom/molecule) with the thermal (as well as coherent) radiation field. Such a QED description was given long before by Jaynes and Cummings but for interactions with a single cavity mode in a resonant cavity~\cite{Jaynes}. However, our generalized semiclassical result is still useful for interactions in the free space with the broad-band modes near around the resonance frequency. It would be important in relating the Rabi model with Einstein's rate (master) equations which are useful in describing nonequilibrium phenomena in terms of the fundamental processes.

\begin{figure}
\includegraphics[width=0.98 \linewidth]{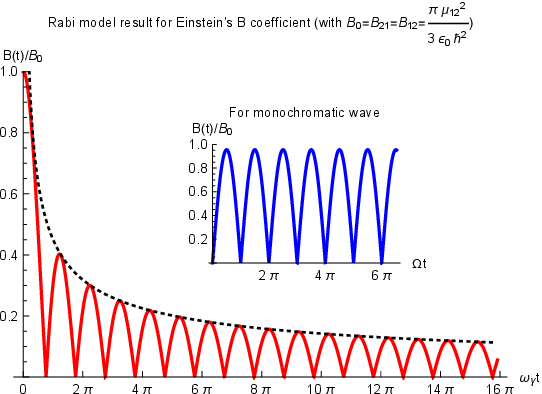}
\caption{Solid line follows Eq.~(\ref{eqn:9}), and represents the Rabi model result for Einstein's $B$ coefficient. Dotted line ($B(t)/B_0\equiv\sqrt{2/\pi\omega_\gamma t}$) represents envelope for the oscillations in the $B$ coefficient. Inset follows Eq.~(\ref{eqn:9b}) and represents the $B$ coefficient for the same system interacting with a monochromatic wave of a single polarization. 
\label{fig1}}
\end{figure}

The time-derivative of the r.h.s. of Eq.~(\ref{eqn:6}) is the transition probability per unit time ($R_{2\rightarrow1}(t)=\frac{d}{d t}P_{2\rightarrow1}(t)$), i.e., the rate of stimulated emission from the state $|\psi_2\rangle$ to the state $|\psi_1\rangle$, which we get from Eq.~(\ref{eqn:6}), as
\begin{eqnarray}\label{eqn:7}
R_{2\rightarrow1}(t)&=&\frac{\mu_{12}^2u(\omega_0)}{3\epsilon_0\hbar^2}\bigg[\int_{-\infty}^{-\omega_\gamma}\frac{\sin(\Omega t)}{\Omega}\frac{1}{\sqrt{1-(\omega_\gamma/\Omega)^2}}d\Omega\nonumber\\&&+\int_{\omega_\gamma}^\infty\frac{\sin(\Omega t)}{\Omega}\frac{1}{\sqrt{1-(\omega_\gamma/\Omega)^2}}d\Omega\bigg]\nonumber\\&=&\frac{\pi\mu_{12}^2u(\omega_0)}{3\epsilon_0\hbar^2}J_0(\omega_\gamma t),
\end{eqnarray}
where $J_0(\omega_\gamma t)$ is the Bessel function of the order zero of the first kind. The rate of absorption ($R_{1\rightarrow2}(t)=\frac{d}{d t}P_{1\rightarrow2}(t)$) from the state $|\psi_1\rangle$ to $|\psi_2\rangle$, on the other hand, is just the opposite, i.e., $R_{1\rightarrow2}(t)=-R_{2\rightarrow1}(t)$ as $P_{2\rightarrow1}(t)+P_{1\rightarrow2}(t)=1$. From Eqs.~(\ref{eqn:3}) and (\ref{eqn:7}), it is clear that the rate of stimulated absorption from the state $|\psi_1\rangle$ to the state $|\psi_2\rangle$ would be
\begin{eqnarray}\label{eqn:8}
R_{1\rightarrow2}(t)=-R_{2\rightarrow1}(t)=-\frac{\pi\mu_{12}^2u(\omega_0)}{3\epsilon_0\hbar^2}J_0(\omega_\gamma t).
\end{eqnarray}
Einstein's $B$ coefficient was already defined as the rate of stimulated transitions (emission or absorption) per unit energy density of radiations per unit (angular) frequency interval around the Bohr frequency. However, the rate of transitions in Eq.~(\ref{eqn:8}) quasi-periodically alters its sign in the course of time as if the absorption becomes emission and vice-versa whenever there is an alternation of the sign. Thus, following the definition of the $B$ coefficient, the feature of alternating the sign of the transition rates and the fact $R_{1\rightarrow2}(t)=-R_{2\rightarrow1}(t)$, we get Einstein's $B$ coefficient ($B(t)=|R_{1\rightarrow2}(t)|/u(\omega_0)=|R_{2\rightarrow1}(t)|/u(\omega_0)$) from Eqs.~(\ref{eqn:7}) and (\ref{eqn:8}), as
\begin{eqnarray}\label{eqn:9}
B(t)=B_0|J_0(\omega_\gamma t)|,
\end{eqnarray}
where $B_0=\frac{\pi\mu_{12}^2}{3\epsilon_0\hbar^2}$ is the original $B$ coefficient obtained by Dirac~\cite{Dirac,Griffiths}. The $B$ coefficient would be unaltered if we alter the initial conditions, by taking $c_1(0)=1$ and $c_2(0)=0$, as we always have $R_{1\rightarrow2}(t)=-R_{2\rightarrow1}(t)$. We show the time-dependence in the $B$ coefficient in FIG. \ref{fig1}. We are not able to compare this result with the existing experimental data because they have not been obtained by any direct measurement (as far as we know); rather, experimentalists apply time-dependent perturbation theory for the indirect measurement of the $B$ coefficient~\cite{Lawrence}.  

\subsection{Effect of quantum fluctuations and thermal fluctuations  in a resonant cavity}
The vacuum Rabi flopping of the two-level system ($^{87}$Rb), however, was observed by Brune \textit{et al.} in a high-$Q$ open resonant (Fabry--Perot) cavity at a very low temperature~\cite{Brune}. Two circular mirrors of diameter $50$mm each were kept nearly $27$mm apart in the cavity set-up so that the $9$th harmonic among all the modes of the standing waves in the cavity-geometry corresponds to the Bohr frequency ($\omega_0=2\pi\times51.099\times10^9$ Hz) for the circular Rydberg states $|n=50\rangle$ and $|n=51\rangle$ of the $^{87}$Rb atom~\cite{Brune}. Their observation of the vacuum Rabi flopping among these two states was phenomenologically well explained by Wilczewski and Czachor by extending the Jaynes-Cummings theory for the resonant cavity~\cite{Wilczewski}. Wilczewski and Czachor considered a number of fitting parameters and fitting functions (e.g., those in the transition probabilities expressed in Eq.~(29) or (30) or (31) in their paper) for explaining the vacuum Rabi flopping by considering the rates of the stimulated transitions to be time-independent. In contrast, no fitting functions and fitting parameters have been used in deriving our free-space transition probability (Eq.~(\ref{eqn:6b})). In what follows, we are now going to derive an expression for the cavity-space transition probability in a similar way by further considering natural decay and the Ohmic losses from the high-$Q$ cavity. 

Losses of the total stored average electromagnetic field energy ($\mathcal{U}(t)$) from the resonant cavity ($\omega\rightarrow\omega_0$) at the constant rate $\omega_0/Q$ {for unit $\mathcal{U}(t)$} results the time-dependence: $\mathcal{U}(t)=\mathcal{U}(0)\text{e}^{-\omega_0t/Q}$. This leads to the Lorentzian broadening of the average energy density per unit (angular) frequency interval in the frequency domain, as $u_c(\omega)=u(\omega_0)\frac{(\omega_0/Q)^2}{4(\omega-\omega_0)^2+(\omega_0/Q)^2}$ where $Q$ is the mode quality factor of the resonant cavity~\cite{Qfactor}. A form similar to the Lorentzian distribution also appears for the spontaneous emission from the upper level to the lower one with the natural decay rate $\gamma=A$ as $u_d(\omega)=u(\omega_0)\frac{A^2}{4(\omega-\omega_0)^2+A^2}$~\cite{Weisskopf}. {These two line-shapes ($u_c(\omega)$ and $u_d(\omega)$) can be convolved}---neglecting Doppler, thermal, and other type of broadening at a low temperature and low number density of atoms/molecules---to write~{\cite{Line-Broadening}}
\begin{eqnarray}\label{eqn:6bb}
u'(\omega)=u(\omega_0)\frac{\Gamma^2}{4(\omega-\omega_0)^2+\Gamma^2},
\end{eqnarray}
where $\Gamma=A+\omega_0/Q$ represents the net decay rate {for the two-level system in the resonant cavity}. Eq.~(\ref{eqn:4}) would result the cavity-space stimulated emission probability if $u(\omega)$ is replaced by $u'(\omega)$ of Eq.~(\ref{eqn:6bb}). Thus, we recast Eq.~(\ref{eqn:4}) with the Lorentzian distribution---in the same manner as we have reached Eq.~(\ref{eqn:6b})---as
\begin{eqnarray}\label{eqn:6c}
P_{2\rightarrow1}(t)&\simeq&\frac{2\omega_\gamma(1+2\omega_\gamma/ \Gamma)}{\pi}\nonumber\times\\&&\int_{\omega_\gamma}^\infty\bigg[\frac{\Gamma^2}{4(\Omega^2-\omega_\gamma^2)+\Gamma^2}\bigg]\frac{\sin^2(\Omega t/2)}{\Omega\sqrt{\Omega^2-\omega_\gamma^2}}\text{d}\Omega~~~~~
\end{eqnarray}
where the pre-factor takes care of the normalization of the transition probability, the factor $2$ in the numerator takes care of the integration in the domain ($-\infty,-\omega_\gamma$], and the Rabi flopping frequency is now modified to
\begin{eqnarray}\label{eqn:6d}
\omega_\gamma=\frac{2\pi\mu_{12}^2u(\omega_0)}{3\epsilon_0\hbar^2}\frac{1}{1+2\omega_\gamma/\Gamma}.
\end{eqnarray}
Here no contributions of the thermal photons (i.e., $u_T(\omega_0)=0$ and $u(\omega_0)=u_q(\omega_0)=\frac{\hbar\omega_0^3}{2\pi^2c^3}$) correspond to the vacuum Rabi flopping (with frequency $\omega_\gamma=[-1+\sqrt{1+4\frac{\mu_{12}^2\omega_0^3}{3\pi c^3\epsilon_0\hbar}\frac{2Q}{\omega_0+AQ}}]/\frac{4Q}{\omega_0+AQ}$) at the resonance, and further sending the net decay rate ($\Gamma=A+\omega_0/Q$)  to $\infty$ (i.e., sending either $Q$ to $0$ or $A$ to $\infty$) corresponds to the vacuum Rabi flopping (with frequency $\omega_\gamma=\frac{\mu_{12}^2\omega_0^3}{3\pi c^3\epsilon_0\hbar}$) in free space. The form of the later exactly matches with that of the result of the Jaynes-Cummings theory obtained even for a single cavity mode~\cite{Jaynes}. It is easy to conclude from Eqs.~(\ref{eqn:6c}) and (\ref{eqn:6d}) that, the transition probability takes on the expression $P_{2\rightarrow1}(t)=\sin^2(\omega_\gamma t/2)$ for $A\rightarrow0$ and $Q\rightarrow\infty$ (or $\triangle\omega\rightarrow0$). The vacuum Rabi flopping frequency vanishes in this limit, as the two-level system does not find even a single mode to couple with in the loss-less resonant cavity without any {natural decays}. 

We represent the r.h.s. of Eq.~(\ref{eqn:6c}) by the dashed line in FIG. \ref{fig1b} for the reported experimental value of $Q$-factor ($Q=7\times10^7$) and the calculated value of the $A$ coefficient (\footnote{Eq.~(\ref{eqn:6d}) has been used to find out $\mu_{12}$, which has been used in determining the value of $A=\frac{\omega_0^3\mu_{12}^2}{3\pi\epsilon_0\hbar c^3}$~\cite{Weisskopf,Griffiths} in free space for the relevant parameters used in the experiment in the open resonant cavity~\cite{Brune}.} $A=0.5536116\times10^6$/s) for the spontaneous emission from the circular Rydberg state $|n=51\rangle$ to $|n=50\rangle$ of $^{87}$Rb atoms in the open resonant cavity of geometrical volume $V=\pi(50/2)^2\times27$ mm$^3$~\cite{Brune}. Substantial deviation of the solid line (Eq.~(\ref{eqn:6b}) for the stimulated emission) from the dashed one or the experimental data mainly comes from the natural decay in the open resonant cavity. Natural decay is usually enhanced in the high-$Q$ ($Q\gg1$) resonant cavity with respect to that in free space by the Purcell factor, $\frac{3(2\pi c/\omega_0)^3Q}{4\pi^2V'}$~\cite{Purcell}, where $V'$ is the mode-volume of the cavity. Here $V'$ ideally goes to $\infty$ for the open cavity. Such an enhancement effect (called the Purcell effect) is not applicable directly to the system and the cavity of our interest, as although the $Q$-factor ($Q=7\times10^7$) is high, the ratio of the free-space decay rate $A=0.5536116\times10^6$/s to the Bohr frequency ($\omega_0=2\pi\times51.099\times10^9$ Hz) is not negligible enough to contain a single bound mode of the electromagnetic field in the open resonant cavity.
	
However, we anticipate  replacement of the $Q$-factor by $Q/(1+AQ/\omega_0)$~\footnote{Net quality factor corresponding to the net decay rate ($\Gamma=A+\omega_0/Q$) would be $Q'$ such that $\Gamma=\omega_0/Q'$. Thus, we have $Q'=Q/(1+AQ/\omega_0)$.} in the Purcell factor to capture the Purcell effect substantially in such a cavity with further replacement of the mode-volume by an effective finite mode-volume ($V_{eff}$). The effective mode-volume would considerably increase the value of the $A$ coefficient from $A=0.5536116\times10^6$/s to $A=1\times10^6$/s for fitting Eq.~(\ref{eqn:6c}) with the experimental data~\cite{Brune} for the effective mode-volume $300.7$ times the geometrical volume of the cavity. We show the fitting of Eq.~(\ref{eqn:6c}) by the dotted line in FIG. \ref{fig1b} for the enhanced decay in the open resonant cavity. Natural decay from the lower level ($|n=50\rangle$) is ignored in our analysis as radiative lifetime ($30$ ms~\cite{Brune}) is quite longer than the time scale involved in FIG. \ref{fig1b}.  However, agreement of our result (Eq.~(\ref{eqn:6c}); dashed line, FIG. \ref{fig1b}) with the experimental data~\cite{Brune} gives us enough confidence to go ahead with the generalized semiclassical model to show the time-dependence in the rate of stimulated transitions and its natural consequences for nonequilibrium statistical mechanics of a two-level system exposed to the thermal radiation field in free space at any temperature.

While the rate of stimulated emission can be found to be primarily controlled by the convolved line-shape in Eq.~(\ref{eqn:6bb}) as $\frac{d P_{2\rightarrow1}(t)}{dt}\propto \text{e}^{-\Gamma t}$ in short time scale ($\omega_\gamma t\ll1$),  the rate would be found to be primarily controlled, as shown in the next section, by the oscillatory part in Eq.~(\ref{eqn:6c}) in the long time scale ($\omega_\gamma t\gg1$). Thus the line shapes, even if  heuristically structured, have no major roles in the long time behaviour of the population dynamics which would be specially important if we analyse the situation near the thermodynamic equilibrium of the system in the weak coupling regime ($\omega_\gamma/A\ll1$).

%%%%%%%%%%%%%%%%%%%%%%%
\section{Einstein's rate equations and their solutions for a two-level system in thermal radiation field in free space}
\begin{figure}
\includegraphics[width=0.98 \linewidth]{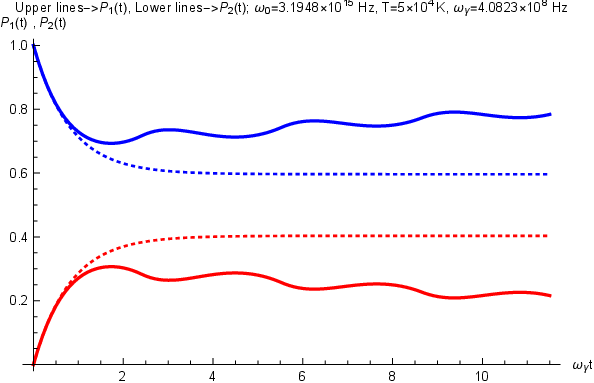}
\caption{
Occupation probabilities for the $3s_{\frac{1}{2}}$ and $3p_{\frac{1}{2}}$ states of an $^{23}$Na atom in the thermal radiation field with the condition that the system initially was in the lower level. Upper and lower solid lines follow Eqs.~(\ref{eqn:13}) and (\ref{eqn:14}) respectively for the parameters as mentioned in the figure corresponding to $\mu_{12}=2.5ea_0=2.1196\times10^{-29}$Cm~\cite{Boyd}. Lower and upper dotted lines represent Einstein probabilities for the same system, and follow Eq.~(\ref{eqn:12b}) and its follow-up respectively. 
\label{fig2}}
\end{figure}

It is to be noted that $\omega_\gamma$, however small, can greatly influence the time-evolution of the statistical mechanical occupation probabilities $P_1(t)$ and $P_2(t)$ of the states $|\psi_1\rangle$ and $|\psi_2\rangle$ respectively even for the case of thermal radiation field. Time-evolution of the occupation probabilities are to be determined from Einstein's rate (master) equations~\cite{Einstein,Griffiths,Feynman} which are now revised with the time-dependent stimulated transition probabilities, as:
\begin{eqnarray}\label{eqn:10}
\frac{d P_2}{dt}=-A P_2(t)-|R_{2\rightarrow1}(t)|P_2(t)+|R_{1\rightarrow2}(t)|P_1(t)
\end{eqnarray}
and
\begin{eqnarray}\label{eqn:11}
\frac{d P_1}{dt}=A P_2(t)+|R_{2\rightarrow1}(t)|P_2(t)-|R_{1\rightarrow2}(t)|P_1(t)
\end{eqnarray}
where $A=\frac{\omega_0^3\mu_{12}^2}{3\pi\epsilon_0\hbar c^3}\ge0$~\cite{Weisskopf,Griffiths} is the original (time-independent) Einstein's $A$ coefficient which represents the rate of spontaneous emission from the upper level to the lower level due to the quantum (vacuum) fluctuations. The time-evolution of these probabilities, because of the constraint $P_1(t)+P_2(t)=1$, can be solely determined from any one of the above two equations, say Eq.~(\ref{eqn:10}), with $P_1(t)$ replaced by $1-P_2(t)$. Thus, we recast Eq.~(\ref{eqn:10}) with the time-dependent rate of absorption, $R(t)=R_{1\rightarrow2}(t)=-R_{2\rightarrow1}(t)$, as
\begin{eqnarray}\label{eqn:12a}
\frac{d P_2}{dt}=|R(t)|-(A+2|R(t)|)P_2(t).
\end{eqnarray}
The rate of stimulated transitions was found in ref.~\cite{Dirac} to be time-independent: $|R(t)|=|R(0)|=B_{12}u(\omega_0)=B_{21}u(\omega_0)\ge0$, within the first order time-dependent perturbation theory of quantum mechanics with $u(\omega_0)=u_T(\omega_0)$~\cite{Griffiths}. Weisskopf and Wigner determined the rate coefficient $A$ within the domain of quantum field theory~\cite{Dirac,Weisskopf}. Eq.~(\ref{eqn:12a}), in such a case, has a physical solution with the initial condition: $P_2(0)=0$, as~\cite{Einstein,Griffiths}
\begin{eqnarray}\label{eqn:12b}
P_2(t)=\frac{|R(0)|}{A+2|R(0)|}[1-e^{-[A+2|R(0)|]t}]
\end{eqnarray}
which is often equated with the (time-independent) Boltzmann probability, $P_2(\infty)=\frac{\text{e}^{-E_2/k_BT}}{\text{e}^{-E_1/k_BT}+\text{e}^{-E_2/k_BT}}$, in thermodynamic equilibrium for $t\rightarrow\infty$~\cite{Einstein,Griffiths}. Occupation probability of the lower level, on the other hand, can be given by $P_1(t)=1-P_2(t)$. Eq.~(\ref{eqn:12b}) is Einstein's semiclassical result for the occupation probability. Let us call the time-dependent probabilities, $P_1(t)$ and $P_2(t)$, which follow from Eq.~(\ref{eqn:12b}), as Einstein probabilities. Dotted lines in FIGs. \ref{fig2} and \ref{fig2a} represent the Einstein probabilities. Our aim for the rest of the paper is to modify the Einstein probabilities due to the presence of the Rabi flopping in the same system within the generalized semiclassical description. 

We solve Eq.~(\ref{eqn:12a}) with the initial condition, $P_2(0)=0$, for $R(t)$ in Eq.~(\ref{eqn:8}), as
\begin{eqnarray}\label{eqn:13}
P_2(t)&=&|R(0)|\text{e}^{-At-2|R(0)|f_{\omega_\gamma}(t)}\nonumber\\&&\times\int_0^t\text{e}^{A\tau+2|R(0)|f_{\omega_\gamma}(\tau)}|J_0(\omega_\gamma\tau)|\text{d}\tau
\end{eqnarray}
where $f_{\omega_\gamma}(t)$ is given by
\begin{eqnarray}\label{eqn:13x}
f_{\omega_\gamma}(t)&=& _1F_2\bigg(\{\frac{1}{2}\}, \{1,\frac{3}{2}\}, -\frac{\omega_\gamma^2 t^2}{4}\bigg)\big[2\text{U}(J_0(\omega_\gamma t))-1\big]t\nonumber\\&&-\sum_{j=1}^{\floor*{\omega_\gamma t}}\bigg[(-1)^j{\gamma_{0,j}} _1F_2\bigg(\{\frac{1}{2}\}, \{1,\frac{3}{2}\}, -\frac{\gamma_{0,j}^2}{4}\bigg)\nonumber\\&&\times\text{U}(\omega_\gamma t-\gamma_{0,j})\bigg]\frac{2}{\omega_\gamma}
\end{eqnarray}
where $\gamma_{0,j}$ is the $j$th zero of $J_0$ and $\text{U}$ is the unit step function. Now, we get the occupation probability of the lower level from Eq.~(\ref{eqn:13}), as
\begin{eqnarray}\label{eqn:14}
P_1(t)=1-P_2(t).
\end{eqnarray}
Eqs.~(\ref{eqn:13}) and (\ref{eqn:14}) are our generalized semiclassical results for the occupation probabilities of the two states of the two-level system in the thermal radiation field. We plot these probabilities in FIG. \ref{fig2} for the relevant values of the parameters for the $3s_{\frac{1}{2}}$ and $3p_{\frac{1}{2}}$ states of an $^{23}$Na atom. For this plot, we have purposefully considered the temperature to be very high ($T=5\times10^4$ K) so that both the rates of spontaneous ones ($A/\omega_\gamma=0.2393$) and stimulated ones ($|R(0)|/\omega_\gamma=1/2$) are comparable to the Rabi flopping frequency to show oscillations in the occupation probabilities. An $^{23}$Na atom is not expected to be ionized even in such a high temperature, as its first ionization potential is $5.1~$eV=$59183~k_B$K. While the occupation probability ($P_2(t)$) of the upper level asymptotically (i.e., for $\omega_\gamma t\gg1$) vanishes as $\frac{|R(0)|}{A}\sqrt{\frac{2}{\pi\omega_\gamma t}}$, the occupation probability ($P_1(t)$) of the lower level asymptotically reaches unity as $1-\frac{|R(0)|}{A}\sqrt{\frac{2}{\pi\omega_\gamma t}}$. It is clear from FIG.~\ref{fig2} that the occupation probabilities of the two-level system are significantly deviating from the Einstein probabilities (as well as the Boltzmann probabilities) as time evolves, and the system goes away from thermodynamic equilibrium as a consequence of the Rabi flopping with non-zero frequency. Our results, of course, match with Einstein probabilities if Rabi flopping is turned off, i.e., if $\omega_\gamma\rightarrow0$.

\section{The case of monochromatic radiation field}
For the case of a monochromatic light (having a single polarization direction perpendicular to a fixed direction of propagation and time averaged~\footnote{Here, time averaging is taken in the very short time scale of $1/\omega_0$.} energy density $u=\frac{1}{2}\epsilon_0E_0^2$) incident on the two-level system, we need not average over the directions of polarizations as done in Eq.~(\ref{eqn:4}), and can recast Eq.~(\ref{eqn:8}) as
\begin{eqnarray}\label{eqn:9a}
R_{2\rightarrow1}(t)=-R_{1\rightarrow2}(t)=\frac{\mu_{12}^2u}{\epsilon_0\hbar^2}\frac{\sin(\Omega t)}{\Omega}.
\end{eqnarray}
Here, we are not considering the effect of vacuum fluctuations so as to restrict our considerations only to monochromatic field. Of course, it needs to be included if one wants to explain the experimental observations dealing with injected laser light~\cite{Brune,Meekhof}. Now, if we define the $B$ coefficient ($B(t)=|R_{1\rightarrow2}(t)|/[u/\Omega]=|R_{2\rightarrow1}(t)|/[u/\Omega]$) for a monochromatic wave, as the rate of the stimulated transitions (emission and absorption) per unit time average energy density per unit generalized Rabi flopping frequency, then it would be
\begin{eqnarray}\label{eqn:9b}
B(t)=\frac{3B_0}{\pi}|\sin(\Omega t)|.
\end{eqnarray}
We show the time-dependence in the $B$ coefficient in FIG. \ref{fig1} (inset). 

\begin{figure}
\includegraphics[width=0.98 \linewidth]{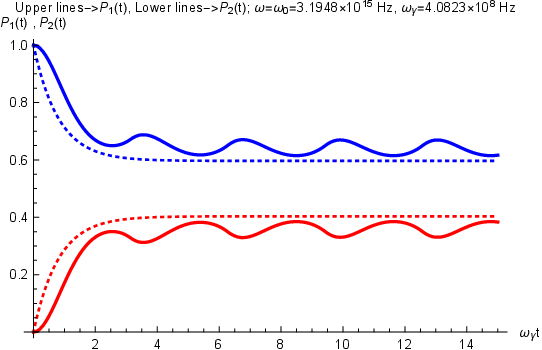}
\caption{
Lower and upper solid lines represent occupation probabilities, and follow Eq.~(\ref{eqn:13a}) and its follow-up for the same parameters of the two-level system at the resonance in the monochromatic radiation field as mentioned in FIG. \ref{fig2}. Adjacent dotted lines represent corresponding Einstein probabilities, and follow Eq.~(\ref{eqn:12b}) and its follow-up respectively. 
\label{fig2a}}
\end{figure}

On the other hand, for the case of the monochromatic wave, we solve Eq.~(\ref{eqn:12a}) with the initial condition $P_2(0)=0$ for $R(t)$ in Eq.~(\ref{eqn:9a}), as
\begin{eqnarray}\label{eqn:13a}
P_2(t)&=&|R(0)|\text{e}^{-At-2|R(0)|g_{\omega_\gamma}(t)}\nonumber\\&&\times\int_0^t\text{e}^{A\tau+2|R(0)|g_{\omega_\gamma}(\tau)}|\sin(\omega_\gamma\tau)|\text{d}\tau
\end{eqnarray}
where $g_{\omega_\gamma}(t)$ is given by
\begin{eqnarray}\label{eqn:13y}
g_{\omega_\gamma}(t)&=&\frac{1 - \cos(\omega_\gamma t)}{\omega_\gamma} [2\text{U}(\sin(\omega_\gamma t))-1]-\frac{2}{\omega_\gamma}\nonumber\\&&\times\sum_{j=1}^{\floor*{\omega_\gamma t}}(-1)^j[1-\cos(j\pi)]\text{U}(\omega_\gamma t-j\pi).
\end{eqnarray}
For this case too, we have $P_1(t)=1-P_2(t)$. The occupation probabilities are quantum mechanical (not statistical mechanical) for the study of the single frequency in the monochromatic wave. We plot these quantum mechanical probabilities in FIG. \ref{fig2a} for the relevant values of the parameters for the same system. It is clear from this figure that, the quantum mechanical probabilities oscillate near the corresponding Einstein probabilities without decay of their amplitudes. Thus, the two-level system (atom/molecule) neither in thermal radiation field nor in the monochromatic radiation field equilibrate with the surroundings as long as the Rabi flopping frequency is non-zero. 

\section{Non-equilibrium statistical mechanical implications}
\begin{figure}
\includegraphics[width=0.98 \linewidth]{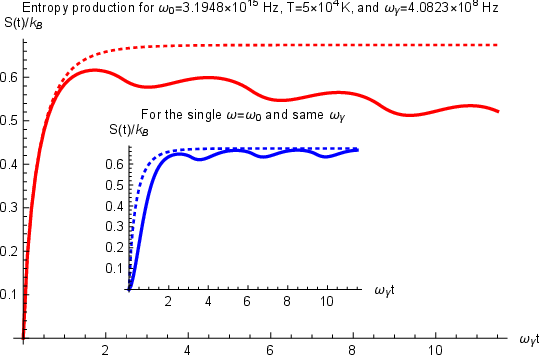}
\caption{Entropy production for the $3s_{\frac{1}{2}}$ and $3p_{\frac{1}{2}}$ states of an $^{23}$Na atom in the thermal radiation field. Plots follow from Eq.~(\ref{eqn:15}) for the parameters as mentioned in FIG.~ \ref{fig2}. Dotted lines represent the same obtained from Einstein probabilities (Eq.~(\ref{eqn:12b}) and its follow-up). 
\label{fig3}}
\end{figure}

Although the two-level system in the rapidly oscillating electromagnetic field makes transitions (if frequency of the oscillations is close to the Bohr frequency of the two levels), the transitions occur over a much larger time scale ($t\sim1/\omega_\gamma$). Thus, it is not exactly known when the system would make a transition. Instead, we know the probability of the transition, and consequently, the occupancy of the two states becomes probabilistic. This loss of information can be quantified by the entropy production of the system. The entropy production of the two-level system either in the thermal radiation field or in the monochromatic radiation field can be written, by following the Pauli--von Neumann formalism of nonequilibrium statistical mechanics, as~\cite{Pauli,Neumann,Feynman}
\begin{eqnarray}\label{eqn:15}
S(t)=-k_B[P_1(t)\ln(P_1(t))+P_2(t)\ln(P_2(t))].
\end{eqnarray}
We illustrate the time-dependence of the entropy in FIG. \ref{fig3} for both the cases; the result corresponding to the monochromatic case is in the inset. Pauli proved the quantum mechanical H-theorem (i.e., $\frac{dS(t)}{dt}\ge0$) even for a single atom/molecule (say, a two-level system) in the radiation field by introducing the Pauli master equation (which is analogous to Einstein's rate equation for $A=0$)~\cite{Pauli,Das}. He considered absolute values of the rates of stimulated transitions to be time-independent for this purpose~\cite{Pauli}. 

However, the two-level system in thermal (or monochromatic) radiation field does not fully evolve spontaneously. The stimulated transitions have control over the evolution of the system specially if the rates of the stimulated transitions are time-dependent. Moreover, the spontaneous emission favour the lower level, as clear from FIG.~\ref{fig2}, once control of the thermal (broad band) radiation to the stimulated transitions is damped (as $\sim1/\sqrt{\omega_\gamma t}$) after sufficiently long time ($\omega_\gamma t\gg1$). Such a damping of the Rabi flopping in free space is caused due to the finite width ($\sim\omega_\gamma$) of the frequency distribution around the resonance as all the frequency components of the thermal radiation field incoherently contribute to the resultant Rabi flopping. Thus, a nonzero finite value of the Rabi flopping frequency causes extraordinary favour on top of the spontaneous transitions to the lower level after sufficiently long time, and consequently, the entropy ($S(t)$) of the two-level system, instead of always increasing with time, asymptotically (i.e., for $\omega_\gamma t\gg1$) vanishes as $k_B\frac{|R(0)|}{A}\sqrt{\frac{2}{\pi\omega_\gamma t}}\big[1-\ln\big[\frac{|R(0)|}{A}\sqrt{\frac{2}{\pi\omega_\gamma t}}\big]\big]$ violating the second law of thermodynamics. This feature is apparent in FIG.~\ref{fig3} where we also have plotted the entropy production based on the Einstein probabilities. Such a damping, however, is not possible for the monochromatic wave, as clear from FIG. \ref{fig2a}, as there is no frequency distribution of the incident waves which causes damping to the Rabi flopping; thus, the Rabi flopping, as clear in the inset of FIG. \ref{fig3}, causes oscillations of the entropy near the non-decreasing semiclassical result (having the saturation value $k_B\ln(2)$). Thus, the second law of thermodynamics is violated for this case too. All the oscillations or the damping are caused for nonzero finite value of the Rabi flopping frequency. Thus, if $\omega_\gamma\rightarrow0$, we again get back Einstein's semiclassical result and validate the second law of thermodynamics for a two-level system (atom/molecule) in the thermal/monochromatic radiation field.

A question naturally arises: whether there would be any change in the occupation probabilities if we take an alternative initial condition such that initially the two-level system (atom/molecule) is at the upper level (i.e., $P_{1}(0)=0$, $P_{2}(0)=1$) like that in Eq.~(\ref{eqn:2}). The Rabi flopping frequency would not certainly change under this alternation. However, some of the results would change, e.g., $P_{2\rightarrow1}(t)$ would be changed to $P_{1\rightarrow2}(t)$ and vice versa, $R(t)$ would be changed to $-R(t)$, an additional term $\text{e}^{-At-2|R(0)|f_{\omega_\gamma}(t)}$ would have to be added to the r.h.s. of  Eq.~(\ref{eqn:13})~\footnote{An additional term $\text{e}^{-At-2|R(0)|g_{\omega_\gamma}(t)}$ would have to be added to the r.h.s. of $P_2(t)$ in Eq.~(\ref{eqn:13a}).}, \textit{etc}. The two solid lines both in FIG. \ref{fig2}  and FIG. \ref{fig2a} would intersect once keeping their individual tails unaltered.

\section{Discussion and Conclusion}
We have shown that the Rabi model result for Einstein's $B$ coefficient depends on time and the Rabi flopping frequency for the two-level system (atom or molecule) in the thermal radiation field at an absolute temperature $T$. This result is accurate for fairly large Bohr frequency ($\omega_0\gg\omega_\gamma$~\footnote{This is also a requirement for the rotating wave approximation, which is inbuilt in the Rabi model, to be valid.}) and fairly higher temperature ($k_BT\gnsim \hbar\omega_\gamma$), and is significantly different from the perturbation result which is not reliable near the resonance in the Rabi flopping. Our analytical result regarding the $B$ coefficient is an invitation for the experimentalists to do direct measurement of the $B$ coefficient. 

Although the limit $\omega_\gamma\rightarrow0$ retrieves the original $B$ coefficient, yet the time-dependence plays a significant role in the population dynamics. The oscillations in the $B$ coefficient, even for very small $\omega_\gamma$, drives the system away from the thermodynamic equilibrium at any finite temperature. This is at odds with the Einstein's assumption about the thermodynamic equilibrium of an atom/molecule with the thermal radiation field~\cite{Einstein}. The predicted equilibrium, however, can be ensured for the case $\omega_\gamma\rightarrow0$, i.e., in absence of the Rabi flopping, as is expected. Nonzero finite value of the light-matter coupling ($\hbar\omega_\gamma$) quasi-periodically drives the two-level system for the multi-frequency modes of the thermal radiation field. We have also obtained results for the same system in the monochromatic radiation field. The drive would be periodic for this case. 

Individual contribution of each frequency component of the thermal radiation field when randomly added damp the Rabi flopping causing extraordinary favour to the lower level on top of the effect of the spontaneous emission. The second law of thermodynamics is not applicable for a driven system. It should be noted in this regard that the consideration of the memory-less transitions, as dealt with by Fermi's golden rule in the time-dependent perturbation theory, is an hypothesis equivalent to the molecular chaos hypothesis which is a necessary condition but not sufficient for reaching equilibrium of a thermodynamically isolated system~\cite{Snoke,Das}. The light-matter coupling ($\hbar\omega_\gamma$) further needs to tend to zero for the system to be not driven by the radiation field, and in turn, to be thermodynamically isolated for observing non-decrease of its entropy as an application of the second law of thermodynamics.

Nonequilibrium quantum statistical mechanics is often modelled with the semiclassical or the quantum master equations which to some extent are generalizations of Einstein's rate equation, such as the Pauli master equation~\cite{Pauli}, the Boltzmann--Uehling--Uhlenbeck equation or  the quantum statistical Boltzmann equation~\cite{Pauli,Das}, the Gorini--Kossakowski--Sudarshan--Lindblad equation~\cite{Lindblad}, the Bloch--Redfield master equation~\cite{Wollfarth,Leggett}, the Caldeira--Leggett master equation~\cite{Caldeira,Leggett}, the quantum Fokker--Planck equation~\cite{Chang}, the adiabatic/nonadiabatic master equation~\cite{Dann}, Van Hove master equation~\cite{VanHove}, and the Nakajima--Zwanzig master equation~\cite{Nakajima}. These equations are either of Markovian master equation or non-Markovian master equation~\footnote{Here only Van Hove and Nakajima-Zwanzig master equations are listed to be of non-Markovian type.} type irrespective of the strength of the system-bath (i.e., light-matter or matter-matter) coupling. None of these equations can be derived fully from either the Schrodinger equation or the Liouville--von Neumann equation or even the Heisenberg equation of motion because the system can not be found in a pure state in the thermal radiation field. These equations (so as Einstein's rate equations) are arrived purely from phenomenological point of view  because (i) the bath is assumed to be not affected by the (much smaller) system, and (ii) the effects of the bath-variables are averaged out with heuristically structured spectral line-shapes of the bath or in turn temporal correlations in the system within various approximations such as the Markov approximation, the Born approximation, \textit{etc.})~\cite{Agarwal2}. The stimulated rate coefficients in the master equations, in general, are time-dependent within the finite time interval after commencement of the light-matter interactions. However, the fundamental processes remain phenomenologically same in both the weak coupling regime and the strong coupling regime even if the (stimulated) rate coefficients are time-dependent (or time-independent) because the bare uncoupled bare states ($|\psi_1\rangle$ and $\psi_2\rangle$) and the energy eigenstates ($\psi_-\rangle$ and $|\psi_+\rangle$), which are dressed after light-matter interactions, belong to the same Hilbert space, and the stimulated transitions take place only between the uncoupled bare states not between the dressed states. Thus, we are generalizing Einstein's rate equations for the time-dependent coefficients and applying to our problem, as the time-dependent coefficients are not bringing any other (new) fundamental processes into the consideration. Application of the generalized rate equations would be a useful model for studying nonequilibrium statistical mechanics for both the weak and the strong light-matter couplings. The generalized rate equations may be classified as a time-dependent semiclassical Markovian master equation as (i) the rates of the stimulated transitions are essentially derived from a semi-classical (Rabi) model and are found to be time-dependent, and (ii) all the occupation probabilities ($P_1$ and $P_2$) in the generalized rate equations are employed at the same time with no memory kernels in the equations. 

While the Rabi flopping usually is studied for strong light-matter interactions ($\omega_\gamma/A\gg1$), Einstein's rate equations are usually applied for weak light-matter interactions ($\omega_\gamma/A\ll1$). Incidentally, the Rabi model, which gives exact results in both the weak coupling regime and the strong coupling regime, is not phenomenologically different from the fundamental processes' point of view. We have been interested in bringing the Rabi flopping and the rate equation together in a single footing for this reason. We have been specially interested in the intermediate regime ($0\lnsim\omega_\gamma/A\lesssim1$) where the partial oscillations, as shown in FIG. (\ref{fig2}) for $\omega_\gamma/A=4.1789$, are expected to be damped for the broadband excitations~\cite{Cohen-Tannudji2}. These partial oscillations, of course, are not periodic~\footnote{This aperiodicity can be linked to the non-regular intervals of the zeros of the Bessel function ($J_0$) in Eq.~(\ref{eqn:7}).} for the nonzero width ($\triangle\omega$) of the frequency band around the resonance. The partial oscillations as shown in FIG. (\ref{fig2a}), however, would neither be damped nor be aperiodic for monochromatic wave, i.e., for extremely narrow band ($\triangle\omega\rightarrow0)$.

Roles of the fundamental processes (the spontaneous emission, the stimulated emission, and the stimulated absorption) in the evolution of the entropy of a system are exemplified by considering the Rabi model as a toy model for the two-level system in the thermal radiation field. The quantum statistical Boltzmann (master) equation is usually employed for time-independent stimulated transition rates. Our time-dependent $B$ coefficient opens a path to go beyond the Pauli--von Neumann formalism of the non-equilibrium statistical mechanics involving the quantum statistical Boltzmann equation~\cite{Pauli,Neumann}. We are, however, not shaking the usual notion of the thermal equilibrium between atoms (or molecules) and black body radiation and Einstein's conclusions, as they are all correct for memory-less transitions under no (external) drives. We are bringing quasi-periodic drive in the calculation of the $B$ coefficient in terms of the Rabi flopping of the two-level system in the thermal radiation field. No system equilibrates under any (external) drives.   

Before concluding the article, we take this opportunity to point out that, our work opens avenues of many interesting research possibilities: (i) how to calculate entropy productions of the laser trapped ultra-cold Bose and Fermi systems by generalizing the toy model, (ii) how to generalize our results for degenerate states of a two-level system, (iii) how to further generalize our generalized semiclassical results, within the purview of the quantum Rabi model (while considering the cavity modes~\cite{Jaynes,Wilczewski}) which has attracted both the experimentalists~\cite{Meekhof,Brune,Haroche} and the theoreticians~\cite{Agarwal,Wilczewski,Xie,Vandaele} alike in the last few decades, and (iii) how to generalize the results for the quasi-continuous splitting of the two levels due to the vibrations in a coupled chain of identical two-level systems.   

\section*{Acknowledgement}
S. Biswas acknowledges partial financial support of the DST, Govt. of India under the INSPIRE Faculty Award Scheme [No. IFA-13 PH-70]. Useful discussions with Prof. J. K. Bhattacharjee (IACS, Kolkata), Prof. Mustansir Barma (TCIS, Hyderabad), Dr. Prasad Perlekar (TCIS, Hyderabad), Prof. S. Dutta Gupta (UoH, Hyderabad), Dr. Ashoka V (UoH, Hyderabad), Dr. Saikat Ghosh (IIT-Kanpur) and Prof. D. S. Ray (IACS, Kolkata) are gratefully acknowledged. {We thank the anonymous reviewer for helping us in significantly improving the presentation of the paper.}

\end{document}